\documentclass{aastex63}

\accepted{to PASP}

\shorttitle{Characterizing Earth-like Exoplanets with JWST}
\shortauthors{Gialluca et al.}

\usepackage{graphicx}
\usepackage{bm}
\usepackage{hyperref}
\usepackage{upgreek}

\begin{document}

\title{Characterizing Atmospheres of Transiting Earth-like Exoplanets Orbiting M Dwarfs with James Webb Space Telescope}

\correspondingauthor{Megan T. Gialluca}
\email{mtg224@nau.edu}

%
\author[0000-0002-2587-0841]{Megan T. Gialluca}
\affiliation{Department of Astronomy and Planetary Science, Northern Arizona University, Box 6010, Flagstaff, AZ 86011, USA}
\affiliation{Habitability, Atmospheres, and Biosignatures Laboratory, Northern Arizona University, Flagstaff, AZ 86011, USA}

\author[0000-0002-3196-414X]{Tyler D. Robinson}
\affiliation{Department of Astronomy and Planetary Science, Northern Arizona University, Box 6010, Flagstaff, AZ 86011, USA}
\affiliation{Habitability, Atmospheres, and Biosignatures Laboratory, Northern Arizona University, Flagstaff, AZ 86011, USA}
\affiliation{NASA Astrobiology Institute’s Virtual Planetary Laboratory, University of Washington, Box 351580, Seattle, WA 98195, USA}

\author{Sarah Rugheimer}
\affiliation{Atmospheric, Oceanic, and Planetary Physics Dept., University of Oxford, Clarendon Laboratory, Parks Road, Oxford OX1 3PU, UK}

\author[0000-0002-2238-5269]{Fabian Wunderlich}
\affiliation{Institute of Planetary Research, German Aerospace Center, Rutherfordstr. 2, 12489 Berlin, Germany}
%

%
\begin{abstract}
\noindent A number of transiting, potentially habitable Earth-sized exoplanets have recently been detected around several nearby M dwarf stars. These worlds represent important targets for atmospheric characterization for the upcoming NASA {\it James Webb Space Telescope}. Given that available time for exoplanet  characterization will be limited, it is critically important to first understand the capabilities and limitations of {\it JWST} when attempting to detect atmospheric constituents for potentially Earth-like worlds orbiting cool stars. Here, we explore coupled climate-chemistry atmospheric models for Earth-like planets orbiting a grid of M dwarf hosts. Using a newly-developed and validated {\it JWST} instrument model\,---\,the {\it JWST} Exoplanet Transit Simulator (JETS)\,---\,we investigate the detectability of key biosignature and habitability indicator gaseous species for a variety of relevant instruments and observing modes. Spectrally-resolved detection scenarios as well as cases where the spectral impact of a given species is integrated across the entire range of an instrument/mode are considered and serve to highlight the importance of considering information gained over an entire observable spectral range. Our results indicate that detectability of gases at individual wavelengths is overly challenging for \textit{JWST} but integrating the spectral impact of a species across the entire wavelength range of an instrument/mode significantly improves requisite detection times. When considering the entire spectral coverage of an instrument/mode, detections of methane, carbon dioxide, oxygen and water at signal-to-noise ratio 5 could be achieved with observations of several tens of transits (or less) for cloud-free Earth-like worlds orbiting mid- to late-type M dwarfs at system distances of up to 10--15~pc. When compared to previous results, requisite exposure times for gas species detection depend on approaches to quantifying the spectral impact of the species as well as underlying photochemical model assumptions. Thus, constraints on atmospheric abundances, even if just upper limits, by {\it JWST} have the potential to further our understanding of terrestrial atmospheric chemistry.


\end{abstract}
%




%
\section{Introduction} \label{sec:intro}
%

The upcoming {\it James Webb Space Telescope} ({\it JWST}) will offer unprecedented observational capabilities across a large range of research areas within astronomy and astrophysics \citep{gardneretal2006}.  For exoplanetary science, {\it JWST} will provide critical opportunities to study transiting worlds across a range of planet sizes and atmospheric states  \citep{seageretal2009,beichmanetal2014,greeneetal2016}.  Already, community efforts are underway to study and understand the most effective approaches to using {\it JWST} to advance our understanding of exoplanet atmospheres and evolution \citep{stevensonetal2016b,beanetal2018}.

Although Earth-like planets orbiting cool stars can face challenges to their habitability \citep{Heller2011,Shields2016}, they are still among some of our most promising targets for follow-up observation.
A number of terrestrial and sub-Neptune exoplanets orbiting nearby cool stars have already been discovered that are well-suited for observation with {\it JWST} \citep[e.g.][]{charbonneauetal2009, gillonetal2017}.  Previous works addressing how effective {\it JWST} might be at characterizing such worlds present a myriad of predictions ranging from pessimistic to optimistic outlooks.  Early studies predicted that {\it JWST} would struggle to detect atmospheric constituents for Earth-like planets orbiting all but the nearest M dwarfs \citep{valentietal2006, demingetal2009}.  More recently, a number of studies argue that {\it JWST} may be able to detect atmospheric constituents for temperate \citep{cowanetal2015, batalhaetal2018, lustigyaegeretal2019, macdonald&cowan2019, pidhorodetskaetal2020} to warm/hot \citep{morleyetal2017, lincowskietal2019, lustigyaegeretal2019} terrestrial planets orbiting M dwarf hosts given a moderate to substantial investment of observing time.

Building on early work in life detection with {\it JWST} \citep{kaltenegger&traub2009}, a limited number of recent studies have emphasized the potential detectability of biosignatures and habitability indicator species \citep[e.g., water vapor and greenhouse gases; for a review, see][]{robinson2018} for Earth-like worlds orbiting cool stellar hosts with {\it JWST}.  \citet{barstow&irwin2016} investigated detections of ozone for habitable, Earth-like variants of TRAPPIST-1~b, c, and d \citep{gillonetal2016,gillonetal2017}.  These authors found that many tens of observed transits would be required given the predicted capabilities of the Near-InfraRed Spectrograph (NIRSpec) and Mid-InfraRed Instrument (MIRI) aboard {\it JWST}.  \citet{krissansentottonetal2018b} investigated CH$_4$-CO$_2$ disequilibrium biosignatures for an early Earth analog TRAPPIST-1~e variant, finding that such signatures could be detected with roughly 10 transit observations with NIRSpec/PRISM.  Recently, \citet{wunderlichetal2019} expanded {\it JWST} biosignature and habitability indicator searches to Earth-like worlds transiting a variety of M dwarf hosts, and highlighted the detectability of specific absorption bands of water vapor, methane, carbon dioxide, and ozone.  Both \citet{komaceketal2020} and \citet{suissaetal2020} highlight how clouds may strongly inhibit atmospheric characterization of transiting habitable water worlds with {\it JWST}, especially when it comes to water vapor detections (a key habitability indicator).  Finally, and while not exclusively {\it JWST}-focused, \citet{tremblayetal2020} used retrieval analyses on simulated transit observations of an Earth-like TRAPPIST-1~e to understand instrument trade spaces for future exoplanet characterization missions.


Here, we build upon previous {\it JWST} transiting exoplanet studies to explore detections of biosignature and habitability indicator species for modern Earth analog worlds orbiting cool stellar hosts.  As in \citet{wunderlichetal2019}, we adopt a grid of host stars that span the range of M dwarfs and place these stars at various distances from the Sun.  Not only do we investigate detections of individual spectral features \citep[as in][]{wunderlichetal2019}, but we also explore how detections can be enhanced by integrating signals over multiple gas absorption bands \citep[as in][who emphasized non-Earth-like atmospheres in the TRAPPIST-1 system]{louie2018,lustigyaegeretal2019}.

In what follows, we begin by describing our adopted grid of M dwarf hosts and their associated Earth-like planetary companions (Section~\ref{sec:atm_models}). We then introduce and validate a new model for computing {\it JWST} exposure times for transiting exoplanets (Section~\ref{sec:inst_model}). Next, we apply this new tool to our grid of Earth-like exoplanets for various {\it JWST} instrument/mode scenarios (Section~\ref{sec:results}). Finally, we discuss our findings as they relate to previous results (Section~\ref{sec:discuss}) and present a brief summary of our findings (Section~\ref{sec:conclusions}).

%
\section{Atmospheric and Transit Spectrum Models} \label{sec:atm_models}
%

Biosignature detections critically depend on the planetary and host stellar environments \citep[e.g., see recent review by][]{meadowsetal2018}.  Not only does the spectral type of the host star impact transit spectral feature strengths due to the dependence of a transit spectrum on the size of the host, but the shape of stellar spectrum incident on the planet has dramatic consequences for atmospheric chemistry and climate \citep{seguraetal2005, huetal2012, shieldsetal2013, domagalgoldmanetal2014,tianetal2014,rugheimer2013spectral}.  As only the coolest stellar hosts are small enough for {\it JWST} to potentially characterize the atmospheres of any associated potentially habitable planets, our analyses highlight a suite of self-consistent, one-dimensional star-planet models across a full range of M~dwarfs.

Planetary atmospheric models are taken from the coupled climate-photochemistry simulations of \citet{rugheimeretal2015}, which are based on the {\tt EXO-P} one-dimensional (vertical) cloud-free radiative-convective terrestrial planetary atmosphere modeling tool \citep{kaltenegger&sasselov2010}.  The {\tt EXO-P} tool is based, in turn, on a one-dimensional planetary atmospheric climate model \citep{kasting&ackerman1986,pavlovetal2000, haqqmisraetal2008, kopparapuetal2013} and a one-dimensional photochemical model \citep{pavlov&kasting2002, seguraetal2005, seguraetal2007}.  Planetary spectra can be generated via incorporated radiative transfer models \citep{kaltenegger&traub2009,traub&stier1976}.

\begin{figure*}
    \centering
    \includegraphics[width=\textwidth]{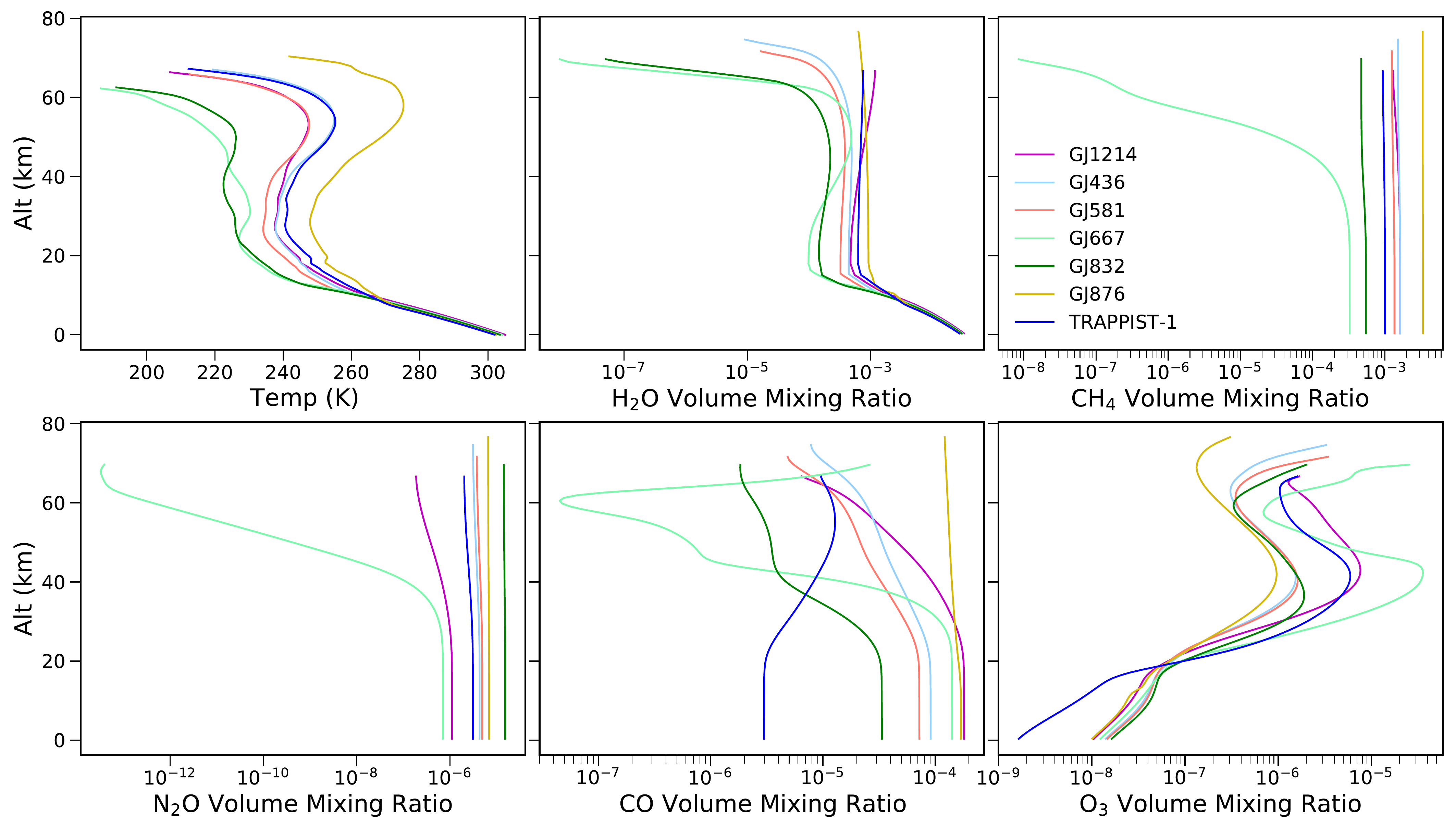}
    \caption{The temperature profiles (top left) and the volume mixing ratios for H$_{2}$O, CH$_{4}$, N$_{2}$O, CO, and O$_{3}$ for the seven different M-type stars considered in our work \citep{rugheimeretal2015}.}
    \label{fig:atmos}
\end{figure*}

Worlds in our adopted {\tt EXO-P} simulations are placed at the 1~AU solar flux equivalent distance (so that the top-of-atmosphere stellar flux for each planet is 1,360~W~m$^{-2}$), are taken to have a radius of $1R_{\oplus}$, and have modern Earth-like biogenic and volcanic gas fluxes.  Our TRAPPIST-1 case is derived from an active M8V simulation in \citet{rugheimeretal2015}, and is most analogous to TRAPPIST-1~d.  While the potential habitability of TRAPPIST-1 d is debated \citep[e.g.,][]{wolf2017}, the present study is simply limited by existing climate-chemistry simulations and should viewed as only potentially representative of how Earth-like worlds in the TRAPPIST-1 system could be characterized with {\it JWST}.  Adopted stellar hosts and associated properties as well as simulated planet orbital distance ($a$) and transit duration ($t_{\rm dur}$) are given in Table~\ref{tab:stars} and are all taken to be consistent with \citet{rugheimeretal2015} and \citet{wunderlichetal2019}. Note that the 1~AU solar flux equivalent distance for these stars is given by,
\begin{equation}
    \frac{a}{{\rm 1~AU}} = \sqrt{\left( \frac{R_{\rm s}}{R_{\odot}} \right)^2 \left( \frac{T_{\rm eff}}{T_{\odot}} \right)^4} \ ,
\end{equation}
where $R_{\rm s}$ is the stellar radius and $T_{\rm eff}$ is the stellar effective temperature. Planetary parameters, including minimum and maximum atmospheric pressure scale height and select volume mixing ratio (VMR) values for water and methane, are shown in Table~\ref{tab:planets}; values in this table are consistent with \citet{rugheimeretal2015}. Note, the mean molecular weight was taken to be Earth-like (29 g~mol$^{-1}$) for all cases. Atmospheric thermal and chemical structures from the {\tt EXO-P} tool are shown in Figure~\ref{fig:atmos}.

For nearly all results, stellar host spectra are taken from \citet{rugheimeretal2015}, which largely rely on high-resolution PHOENIX stellar models \citep{allard2014}. The exceptions are for cases where we compare noise model results to those of \citet{wunderlichetal2019} or the official {\it JWST} exposure time calculator \citep{pontoppidan2016}. Here\,---\,either for consistency with prior results or to mitigate against file upload size limits\,---\,we adopt moderate-high resolution stellar spectra from \citet{france2016muscles} \citep[and][in the case of TRAPPIST-1]{peacock2019predicting}. Note that the resolving power of the spectra developed in \citet{rugheimeretal2015} are comparable to those presented in \citet{Husser_2013} (i.e.,~generally exceeding 200000). More complete details regarding host stellar spectra and planetary models can be found in \citet{rugheimeretal2015}.

\begin{table*}
\begin{tabular}{ c c c c c c }
 Star & Spectral Type & Stellar Radius ($R_{\odot}$) & $T_{\rm eff}$ (K) &  $a$ (AU) & $t_{\rm dur}$ (hr) \\ \hline
 GJ 667C    & M1.5V & 0.348 & 3350  & 0.117 & 4.50 \\
 GJ 832     & M2V   & 0.480 & 3620  & 0.188 & 4.66 \\
 GJ 436     & M2.5V & 0.455 & 3320  & 0.150 & 3.83 \\
 GJ 581     & M3V   & 0.299 & 3500  & 0.110 & 2.57 \\
 GJ 876     & M4V   & 0.376 & 3130  & 0.110 & 2.85 \\
 GJ 1214    & M4.5V & 0.211 & 3250  & 0.0667 & 1.83 \\
 TRAPPIST-1 & M8V   & 0.121 & 2510  & 0.0228 & 0.89 \\
\end{tabular}
\caption{Stellar parameters, 1~AU solar flux equivalent distance, and transit duration for simulated cases. }
\label{tab:stars}
\end{table*}

\begin{table*}
    \centering
    \begin{tabular}{ c c c c c c c c}
        Host Star & Min Scale Height (km) & Max Scale Height (km) & H$_{2}$O VMR at Surface & Max CH$_{4}$ VMR \\ \hline
        GJ 667C & 5.46 & 8.90 & $3.1\times10^{-2}$ & $3.3\times10^{-4}$\\
        GJ 832 & 5.59 & 8.88 & $3.0\times10^{-2}$ & $5.5\times10^{-4}$ \\
        GJ 436 & 6.42 & 8.88 & $3.0\times10^{-2}$ & $1.6\times10^{-3}$ \\
        GJ 581 & 6.22 & 8.88 & $3.0\times10^{-2}$ & $1.4\times10^{-3}$  \\
        GJ 876 & 7.07 & 8.84 & $2.8\times10^{-2}$ & $3.4\times10^{-3}$ \\
        GJ 1214 & 6.05 & 8.92 & $3.3\times10^{-2}$ & $1.6\times10^{-3}$ \\
        TRAPPIST-1 & 6.21 & 8.83 & $2.8\times10^{-2}$ & $1.0\times10^{-3}$ \\
    \end{tabular}
    \caption{Key planetary parameters.}
    \label{tab:planets}
\end{table*}

High-resolution transit spectra are created using the well-validated {\tt scaTran} addition \citep{robinson2017} to the Spectral Mapping Atmospheric Radiative Transfer ({\tt SMART}) model \citep[developed by D.~Crisp;][]{meadows&crisp1996}.  These simulations take the thermal and chemical vertical structures from the {\tt EXO-P} model as inputs, and incorporate the effects of atmospheric absorption along the slant path as well as refraction \citep[which has a modest impact on transit spectra of Earth-like worlds orbiting early M dwarfs, see][their Figure~4]{robinson2018}.  Ro-vibrational opacities for relevant absorbing species (H$_2$O, CO$_2$, O$_3$, N$_2$O, CO, CH$_4$, O$_2$, and CH$_3$Cl) are derived from the HITRAN database \citep{rothmanetal2013, gordonetal2017}, and collision-induced absorption opacities for O$_2$-O$_2$ and O$_2$-N$_2$ are included \citep{baranovetal2004, gordonetal2017}.


%
\section{Instrument Noise Model} \label{sec:inst_model}
%

A number of {\it JWST} noise models across a range of complexities have been described in the literature \citep[e.g.,][]{raueretal2011, vonparisetal2011, hedeltetal2013, greeneetal2016, pontoppidan2016, batalhaetal2017b, wunderlichetal2019, sarkaretal2020}.  In this section, we describe our own moderate-complexity {\it JWST} simulation tool, known as the JWST Exoplant Transit Simulator (JETS), based in large part on the model described by \citet{wunderlichetal2019} \citep[see also][]{wunderlichetal2020}.  We detail our signal-to-noise ratio (SNR) treatments and present validations of our tool against results from \citet{wunderlichetal2019}, which is similar in complexity to our tool and has been cross-validated against the {\tt Pandeia} tool \citep{pontoppidan2016}.

\subsection{Noise Model Description}

We define the wavelength-dependent ``signal'' due an absorbing molecular species (indicated by sub-script `${\rm m}$') over some integration time  during occultation ($\Delta t$) as the difference in the number of in-transit stellar counts for a model with the species included versus a model where the species has been removed, or,
\begin{equation}
    S_{\lambda} = c_{\rm s} \cdot \Delta t \cdot \delta_{\rm m} \ ,
\end{equation}
where $S_{\lambda}$ is the number of counts and $c_{\rm s}$ is the wavelength-dependent stellar count rate.  Here, $\delta_{\rm m}$ is the wavelength-dependent transit depth difference between a spectral model with an absorbing species included versus a model where that species has been removed, computed consistent with Equation 4 of \citet{Deming_2017}.
Note that this is different from the signal defined in \citet{wunderlichetal2019}, where the difference is taken between a model with an atmosphere and an atmosphere-free case in wavelength ranges centered on gas absorption features --- the approach here, that was also used in \citet{lustigyaegeretal2019}, isolates the effects caused by an absorbing species and will, all else being equal, result in longer requisite integration times for gas detections.

Noise in our model is due to photon counting noise in stellar counts, zodiacal light, telescope thermal emission, dark current, and read noise.  Assuming an equal amount of out-of-transit observing time, the noise counts are then given by,
\begin{equation}
    N_{\lambda} = \sqrt{ 2 \Delta t \left( c_{\rm s} + c_{\rm z} + c_{\rm th} + c_{\rm D} + c_{\rm R} \right) } \ ,
\end{equation}
where the count rate terms are associated with the noise sources just mentioned (respectively).  Following \citet{wunderlichetal2019}, the stellar count rate is given by,
\begin{equation}
    c_{\rm s} = A \cdot \mathcal{T} \cdot \Delta \lambda \cdot \frac{\lambda}{hc} \cdot F_{0,\lambda} \cdot \left( \frac{d_0^2}{d^2} \right) \ ,
\end{equation}
where $A$ is the collecting area of the telescope (25.0~m$^{2}$ for {\it JWST}), $\mathcal{T}$ is the wavelength-dependent telescope and instrument throughput (including detector quantum efficiency effects), $\Delta \lambda$ is the wavelength-dependent spectral element width, $F_{0,\lambda}$ is the wavelength-dependent stellar spectral flux density (in energy per unit area per unit time per unit wavelength) normalized to a distance $d_0$, $d$ is the adopted distance to the stellar system, and the factor of $\lambda/hc$ yields the number of photons per unit energy.  Our zodiacal light count rate is taken from \citet[][]{sarkaretal2020} (their Equations~16--18).  The telescope/instrument thermal emission count rate follows \citet{vonparisetal2011} and is given by, 
\begin{equation}
    c_{\rm th} = A \cdot \epsilon_{\rm sys} \cdot \Delta \lambda \cdot B_{\lambda}\left(T_{\rm sys}\right) \cdot \frac{\lambda}{hc} \ ,
\end{equation}
where $\epsilon_{\rm sys}$ is the effective telescope/system emissivity, $B_{\lambda}$ is the Planck function, and $T_{\rm sys}$ is an effective system temperature.  The dark current count rate is taken as,
\begin{equation}
    c_{\rm D} = D_{\rm c} \cdot N_{\rm pix} \ ,
\end{equation}
where $D_{\rm c}$ is the instrument dark current (in counts or electrons per pixel per unit time) and $N_{\rm pix}$ is the wavelength-dependent number of pixels contributing to a given spectral element.  Finally, the read noise count rate is given by,
\begin{equation}
    c_{\rm R} = R_{\rm n} \cdot N_{\rm pix} \cdot {\rm max}\left( \frac{ c_{\rm s}}{f C_{\rm fw}} \right) \ ,
\end{equation}
where $R_{\rm n}$ is the detector read noise (in counts or electrons per pixel per read), $C_{\rm fw}$ is the full-well capacity of a detector pixel, and $f$ is the fraction of the full-well that can be reached before a read would occur.  Thus, $c_{\rm s}/f C_{\rm fw}$ is the read rate and the ${\rm max}$ serves to identify the wavelength with the largest read rate (generally near the peak of the stellar photon emission rate).

With the expressions above, we can define the wavelength-dependent signal-to-noise ratio for species detection as,
\begin{equation}
    \frac{S_{\lambda}}{N_{\lambda}} = \frac{c_{\rm s} \cdot \delta_{\rm m} } {\sqrt{ 2  \left( c_{\rm s} + c_{\rm z} + c_{\rm th} + c_{\rm D} + c_{\rm R} \right) }} \cdot \Delta t^{\frac{1}{2}} \ ,
\label{eqn:snr_res}
\end{equation}
where the number of observed transits, $N_{\rm t}$, required to detect the transit depth difference for species at a given signal-to-noise and wavelength is obtained by setting $\Delta t = N_{\rm t} \cdot t_{\rm dur}$ in the expression above (where $t_{\rm dur}$ is the transit duration) and solving for $N_{\rm t}$. We account for detector read overheads and associate saturation effects in our exposure times according to treatments in \citet{wunderlichetal2019} which, along with a parameterized treatment of the detector point-spread function (that uses the known instrument dispersion), represents a simplified detector model.

Following \citet{louie2018} and \citet{lustigyaegeretal2019}, the signal-to-noise combined over all wavelengths is then,
\begin{equation}
    \frac{S}{N} = \sqrt{\sum_{\lambda} \left( \frac{S_{\lambda}}{N_{\lambda}} \right)^2 } \ ,
\label{eqn:snr_int}
\end{equation}
where a sum symbol indicates summation over spectral elements (i.e., wavelength).  This expression can also be straightforwardly used to find the required number of transits to detect a transit depth difference for a given species across the entire wavelength range of an instrument at a specified signal-to-noise ratio. This treatment serves to emphasize signal at wavelengths with absorber features and minimize noise contributions in wavelengths with little to no absorption impact. In the detectability studies that follow, we adopt a threshold  signal-to-noise ratio of 5 for detection in both wavelength-resolved and wavelength-integrated scenarios. Our required exposure time (or number of transits) results can be converted to other signal-to-noise ratio thresholds by noting that these exposure times will scale with the square of the threshold signal-to-noise ratio.

Our {\it JWST} simulation tool includes all four on-board instruments: the Near-InfraRed Spectrograph \citep[NIRSpec;][]{bagnascoetal2007}, the Near-InfraRed Imager and Slitless Spectrograph \citep[NIRISS;][]{doyonetal2012}, the Near-InfraRed Camera \citep[NIRCam;][]{horner&rieke04}, and the Mid-InfraRed Instrument \citep[MIRI;][]{wrightetal2004}. Each of these instruments is equipped with multiple observing modes and observing options. The NIRSpec instrument operates over 0.6--5.3~$\upmu$m and, in its Bright Object Time-Series (BOTS) mode, has low-resolution (PRISM) as well as medium- and high-resolution options corresponding to resolving powers of roughly 100, 1000, and 2700, respectively. Additionally, when using NIRSpec medium- or high-resolution options, the observer must also select a filter and disperser combination. Here, NIRSpec can only observe a subset of its overall wavelength range capability. At longer near-infrared wavelengths (2.4--5.0~$\upmu$m), the NIRCam instrument can perform spectroscopic observations with two different grisms and four different wide filters. The NIRISS instrument provides slitless spectroscopy options via a wide field mode (0.8--2.2~$\upmu$m, resolving power of 150) and a single object mode (SOSS; 0.6--2.8~$\upmu$m, resolving power of 700). Finally, MIRI offers low-resolution spectroscopy (LRS; 5--12~$\upmu$m) and medium-resolution spectroscopy (MRS; 4.9--28.8~$\upmu$m) with variable resolving powers that span 40--160 and 1,550--3,250, respectively.

Model information on the individual instruments and observing modes were taken from the \href{https://jwst-docs.stsci.edu/}{{\it JWST} User Documentation}~(2016--2020) \nocite{jwstdocs}, and include the wavelength range(s) over which each instrument operates, resolving power, detector dark current, detector read noise, and the detector full-well capacity. The wavelength-dependent telescope and instrument throughputs come from the {\tt Pandeia} \citep{pontoppidan2016} and {\tt PandExo} \citep{batalhaetal2017} tools. Finally, we assume that all spectral elements have two pixels in the wavelength dimension (i.e., Nyquist sampling), and the number of pixels in the orthogonal direction for each instrument are taken from \citet{batalhaetal2017}.

\begin{figure*}
    \centering
    \includegraphics[width=\textwidth]{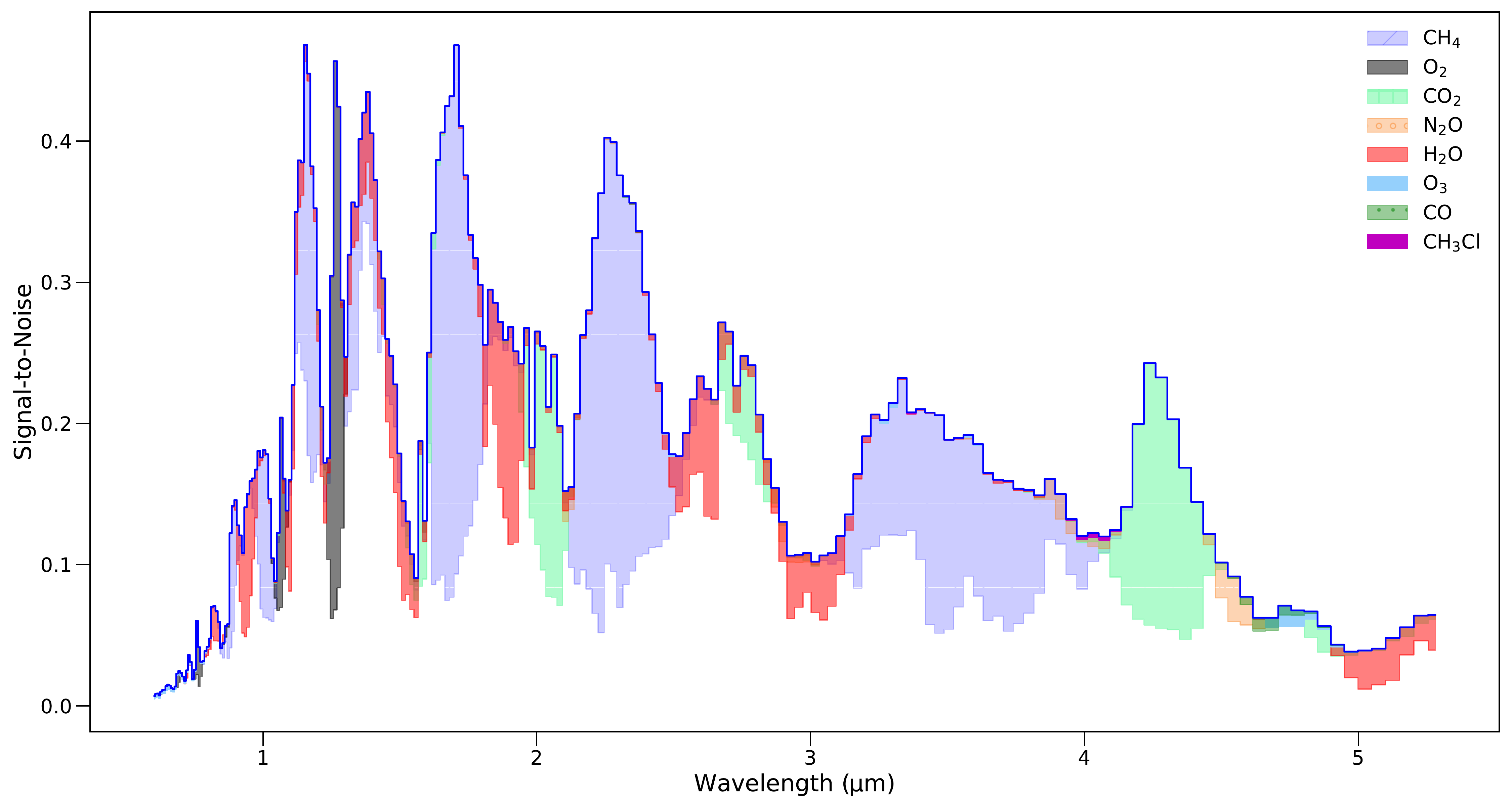} 
    \caption{
         Atmospheric detection signal-to-noise ratio for a hypothetical exo-Earth orbiting TRAPPIST-1 with NIRSpec/PRISM and an integration time of one transit at a fixed distance of 10 pc. Signal-to-noise contributions from key individual species are shown via shading.
    }
    \label{SNRCompare}
\end{figure*}

To demonstrate the utility of our instrument simulator and modeling approach, Figure~\ref{SNRCompare} shows the atmospheric detection signal-to-noise ratio (i.e., the signal-to-noise ratio change when the atmosphere is removed) achieved over a single transit for a hypothetical exo-Earth orbiting TRAPPIST-1. Here, we adopt NIRSpec/PRISM. Contributions from individual species are shown via color shading. Note that signal-to-noise does not drop to zero at a particular wavelength if the strongest absorber at that wavelength is removed due to the smaller contributions of all other gases in that spectral range.

\subsection{Model Validation} \label{sec:validation}
\begin{figure*}
    \centering
    \includegraphics[width=\textwidth]{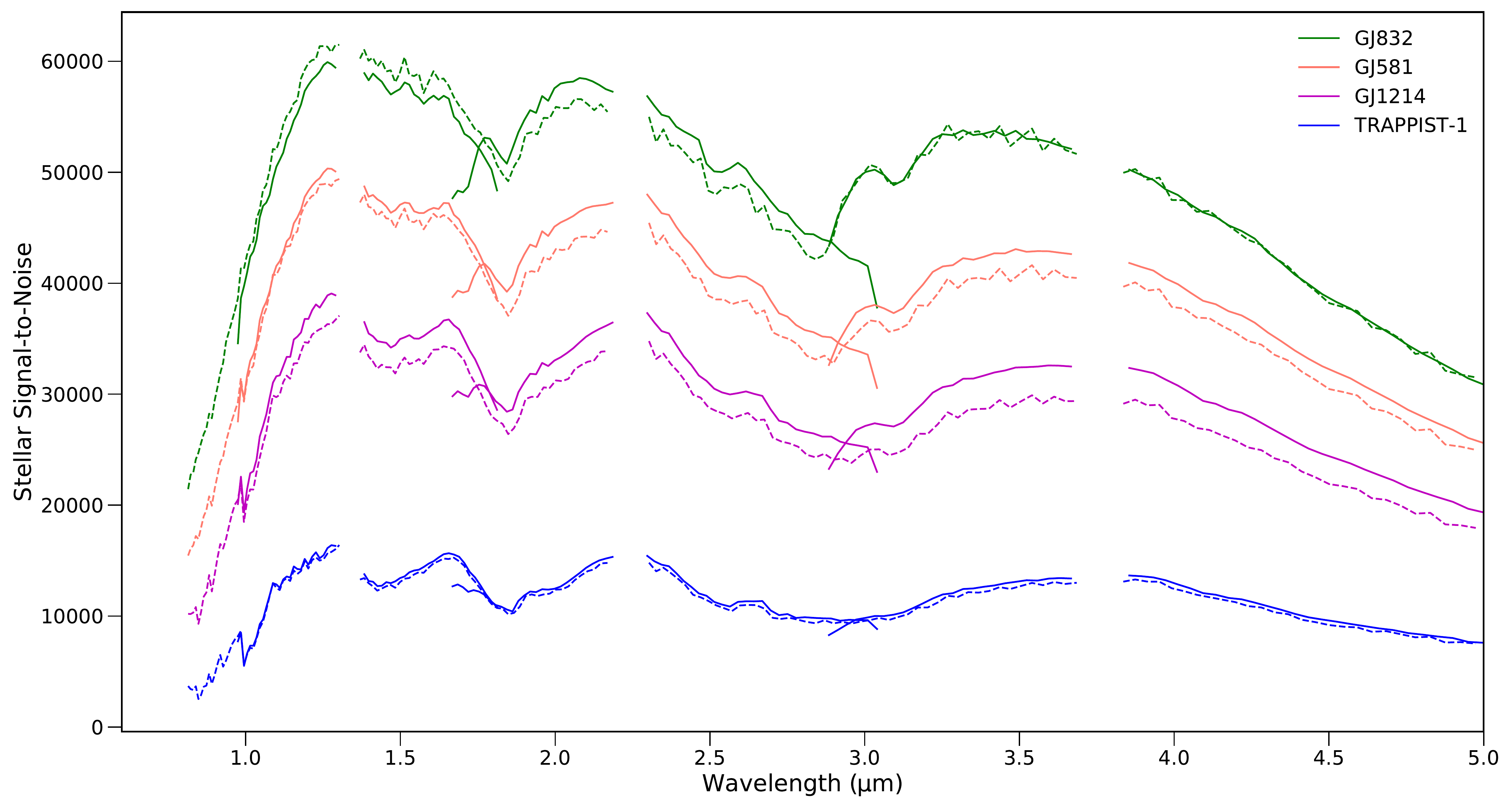} 
    \caption{
        Stellar signal-to-noise ratio for GJ832, GJ581, GJ1214, and TRAPPIST-1, from our {\it JWST} model (solid lines) and from  \citet{wunderlichetal2019} (dashed lines). For this comparison, all stars are placed at 10 pc, a 1~hr integration time is adopted, and all NIRSpec high-resolution modes are binned to a resolving power of 100.
    }
    \label{StellarCompare}
\end{figure*}

We validate our noise model via a comparison to the {\it JWST} Exposure Time Calculator \citep{pontoppidan2016} and to results from \citet{wunderlichetal2019} (which, as mentioned earlier, has been cross-validated against the {\tt Pandeia} tool). Our validation emphasizes NIRSpec as most results shown below are for this instrument. Figure~\ref{StellarCompare} compares the stellar signal-to-noise ratio achieved for TRAPPIST-1, GJ1214, GJ832, and GJ581 in a 1~hr integration with a collection of NIRSpec high-resolution modes from \citet{wunderlichetal2019} and our model. Identical input stellar spectra were adopted \citep{france2016muscles,peacock2019predicting}. Overall, agreement is strong (especially for low temperature stars like TRAPPIST-1) and within the uncertainty range for an instrument with unmeasured ``on-sky'' performance.

\begin{figure*}
    \centering
    \includegraphics[width=\textwidth]{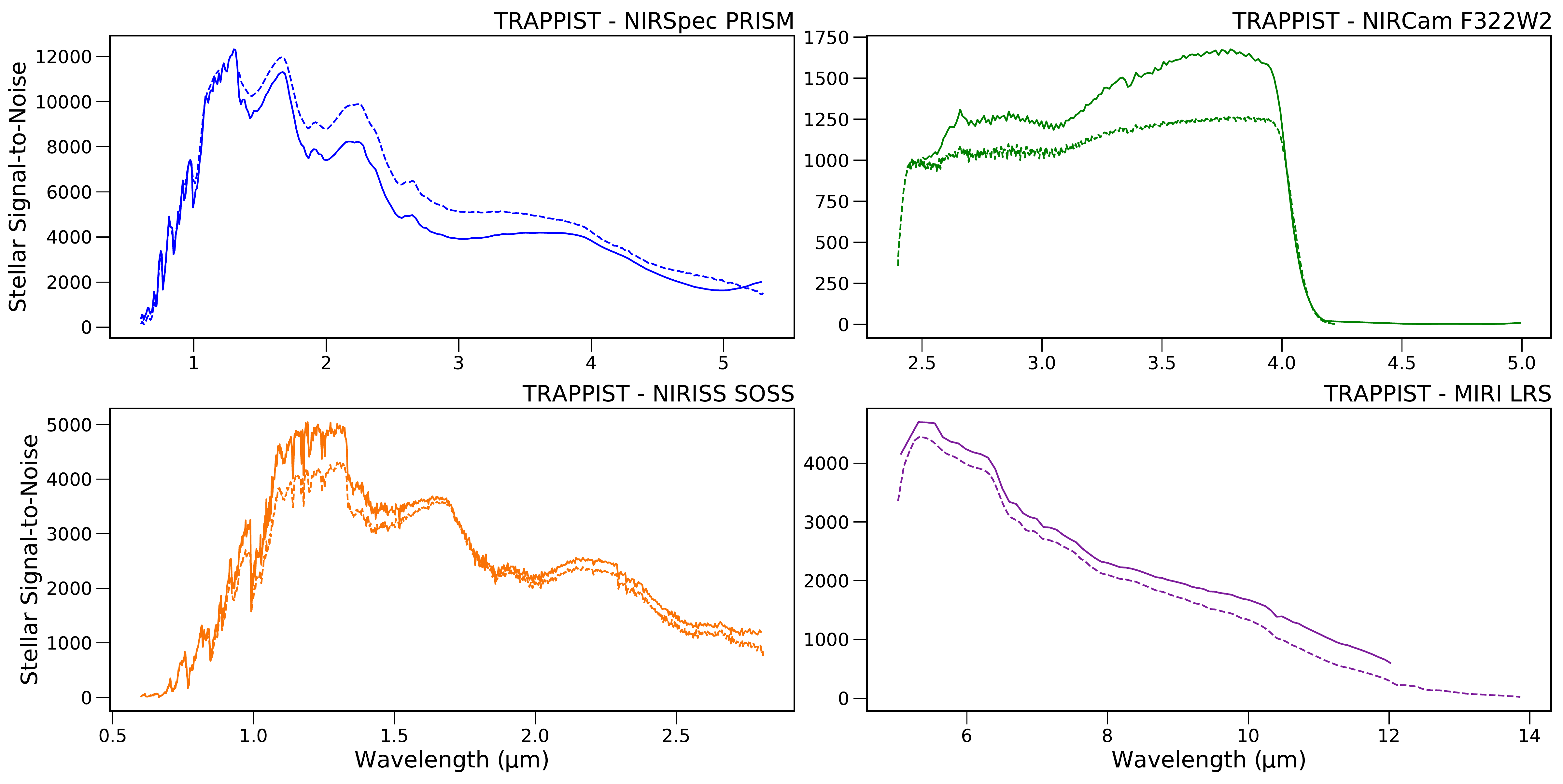}
    \caption{Stellar signal-to-noise ratio for a 1-hour observation of a TRAPPIST-1 analog at 10~pc for NIRSpec/PRISM, NIRCam/F322W2, NIRISS/SOSS, and MIRI low-resolution from our {\it JWST} model (solid lines) and from the offical {\it JWST} Exposure Time Calculator \citep{pontoppidan2016}.}
    \label{fig:ETC_compare}
\end{figure*}

Figure~\ref{fig:ETC_compare} compares the stellar signal-to-noise ratio for a TRAPPIST-1 analog at 10~pc using NIRSpec/PRISM, NIRCam/F322W2, NIRISS/SOSS, and MIRI low-resolution with a 1~hr integration time between our simple noise model and results from the official {\it JWST} Exposure Time Calculator \citep{pontoppidan2016}. Agreement is strong for both NIRISS and MIRI, while our simplified model slightly under-predicts the SNR for NIRSpec. Our simplified noise model over-predicts SNRs for NIRCam by about 20\% in the 2.5--3.0~$\upmu$m range, growing to about 50\% in the 3.5--4.0~$\upmu$m range. However, as is discussed in Section~\ref{sec:discuss}, primary results presented below rely on NIRSpec and MIRI over NIRCam in all cases.

%
\section{Results} \label{sec:results}

Using our combination of atmospheric models (Section~\ref{sec:atm_models}) and simulation tools (Section~\ref{sec:inst_model}), we aim to quantify the potential performance of {\it JWST} for detecting chemical species in the atmospheres of hypothetical Earths transiting a variety of different M dwarf hosts (see Table~\ref{tab:stars}). 
However, the suite of included radiatively-active atmospheric species (8; CH$_4$, CH$_3$Cl, CO, CO$_2$, H$_2$O, N$_2$O, O$_2$, and O$_3$), stellar hosts (7), instruments (4), and observing modes (varies by instrument) makes for an overly-large number of permutations when studying detectability. 
Thus, Section~\ref{sec:InstrumentPerformance} begins by exploring the collection of relevant instrument and observing modes for a limited number of stellar hosts and atmospheric species. 
These results help to identify the most promising instrument and observing modes for characterizing the atmospheres simulated here. 
Section~\ref{sec:NtVWav} uses NIRSpec/PRISM (selected for its performance in Section~\ref{sec:InstrumentPerformance} and its wide wavelength range) to explore detectability at individual wavelengths.
While detectability results at a single wavelength are not promising, integrating gas signatures over the full wavelength range of a given instrument and observing mode can significantly improve detection prospects.
Thus, Section~\ref{sec:NtVDis} uses all the best case modes to explore detectability for all included atmospheric species and stellar hosts for an array of distances (0 to 20 pc) by integrating over wavelength at a particular distance to show only distance dependence.

\subsection{Instrument Mode Comparisons} \label{sec:InstrumentPerformance}

\begin{figure*}
    \centering
    \includegraphics[width=\textwidth]{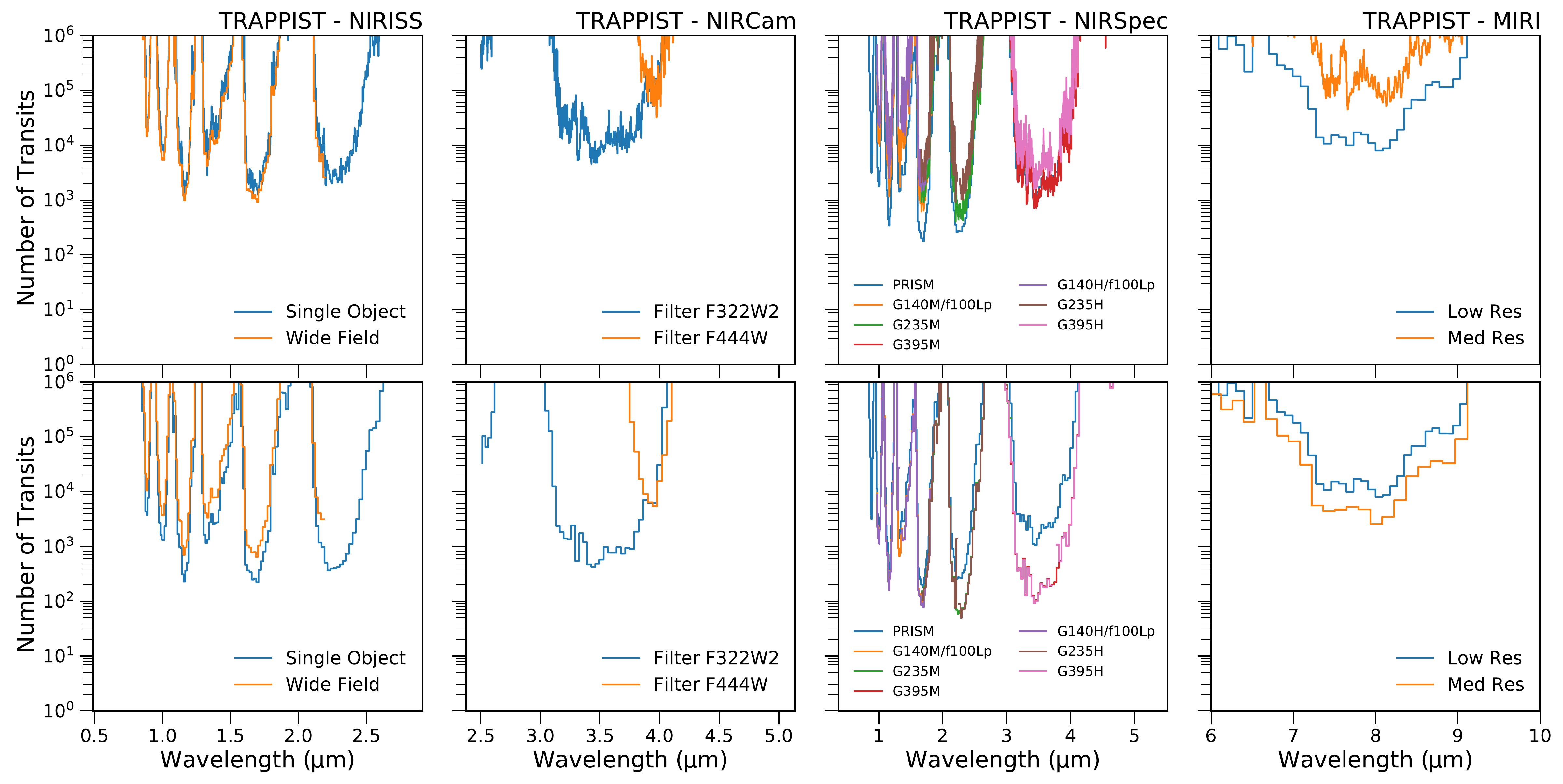} 
    \caption{
        Number of transits required to detect the spectral impact of methane at a signal-to-noise ratio of 5 for a variety of {\it JWST} instruments and modes adopting an Earth-like planet orbiting in the Habitable Zone of a TRAPPIST-1-like host at 10~pc. Top row shows the instruments and observing modes at their native resolution and the bottom row shows the modes binned to a common resolving power of 100 (except for MIRI).
    }
    \label{JWSTPerformanceTrap}
\end{figure*}

\begin{figure*}
    \centering
    \includegraphics[width=\textwidth]{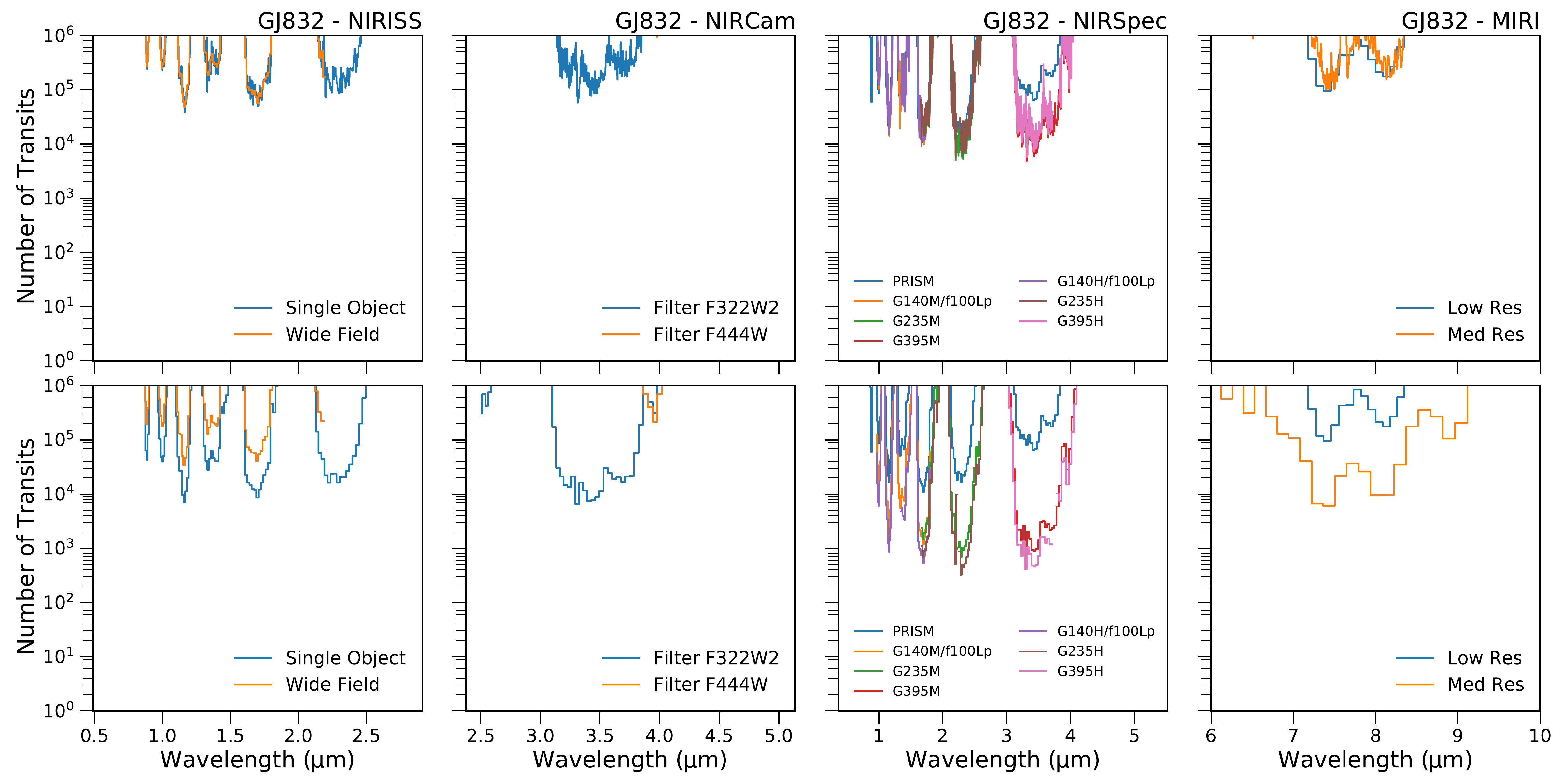} 
    \caption{
        Same as Figure~\ref{JWSTPerformanceTrap} except for a GJ832-like host at 10~pc.
    }
    \label{JWSTPerformance832}
\end{figure*}

Figures~\ref{JWSTPerformanceTrap} and \ref{JWSTPerformance832} show the required number of observed transits to detect the spectrally-resolved (i.e., from Equation~\ref{eqn:snr_res}) impact of methane at a signal-to-noise ratio of 5 for all instruments and relevant observing modes when observing a hypothetical Earth in transit around TRAPPIST-1- and GJ832-like targets at 10~pc. Note that these host stars span the full range of stellar effective temperatures considered here, and we highlight methane as it presents a number of relatively strong features through the near- and mid-infrared. For comparison purposes, the bottom row of Figures~\ref{JWSTPerformanceTrap} and \ref{JWSTPerformance832} shows each observing mode binned to a common resolving power of 100, except for MIRI. As the MIRI low-resolution mode is at coarser resolving power than 100 at some wavelengths, we opted to bin MIRI the medium-resolution mode to be equal to that of low-resolution (i.e., spanning 40--160 across the spectral range). 

Regarding our TRAPPIST-1-like case (Figure~\ref{JWSTPerformanceTrap}), and at native instrument and mode resolutions, results show that NIRSpec/PRISM achieves  methane detection in the smallest number of observed transits. When binned to a common resolving power of 100, both NIRCam/F322W2 and NIRISS/SOSS show comparable performance to NIRSpec/PRISM. However, not considering saturation issues, NIRSpec/PRISM provides greater observing advantages given its broader wavelength coverage as compared to NIRCam/F322W2 and NIRISS/SOSS. Additionally, when binned to a resolving power of 100, NIRSpec medium- and high-resolution modes generally achieve fewer requisite numbers of observed transits as compared to NIRSpec/PRISM. Here, though, it is important to note that since both NIRSpec medium- and high-resolution modes have their wavelength ranges split between three different dispersers, it would take three separate observed transits to achieve an equivalent wavelength coverage as NIRSpec/PRISM. 

Our 10~pc GJ832-like case (Figure \ref{JWSTPerformance832}) demonstrates instrument performances that are, overall, more similar to one another. At native resolution, NIRSpec medium- and high-resolution modes achieve the smallest number of transits to detection. As before, and depending on saturation effects, the broader wavelength coverage of NIRSpec/PRISM may make this the preferred mode over NIRSpec medium- and high-resolution modes.  The NIRCam/F322W2 mode is also comparable to NIRSpec/PRISM at native resolution and exceeds the capabilities of NIRSpec/PRISM when binned to a resolution of 100. Note, once again, that the narrower wavelength coverage of NIRCam/F322W2 may make NIRSpec/PRISM the preferred instrument. Given the results of this subsection, Section \ref{sec:NtVWav} emphasizes NIRSpec/PRISM when exploring the characterization capabilities of {\it JWST}. Section \ref{sec:NtVDis} considers all of NIRSpec/PRISM, NIRSpec high-resolution, NIRCam/F322W2, and NIRISS/SOSS, as they perform similarly (as is shown by Figures \ref{JWSTPerformanceTrap} and \ref{JWSTPerformance832}). Additionally, due to its broad wavelength coverage past 5 microns and its potential for exoplanetary science \citep[see][]{Snellen_2017}, we also consider MIRI medium-resolution as a best case instrument. MIRI medium-resolution was selected over MIRI low-resolution for its slightly better performance in Figures~\ref{JWSTPerformanceTrap} and \ref{JWSTPerformance832}.

\subsection{Resolved NIRSpec/PRISM Gas Detection Times} \label{sec:NtVWav}

\begin{figure*}
    \centering
    \includegraphics[width=\textwidth]{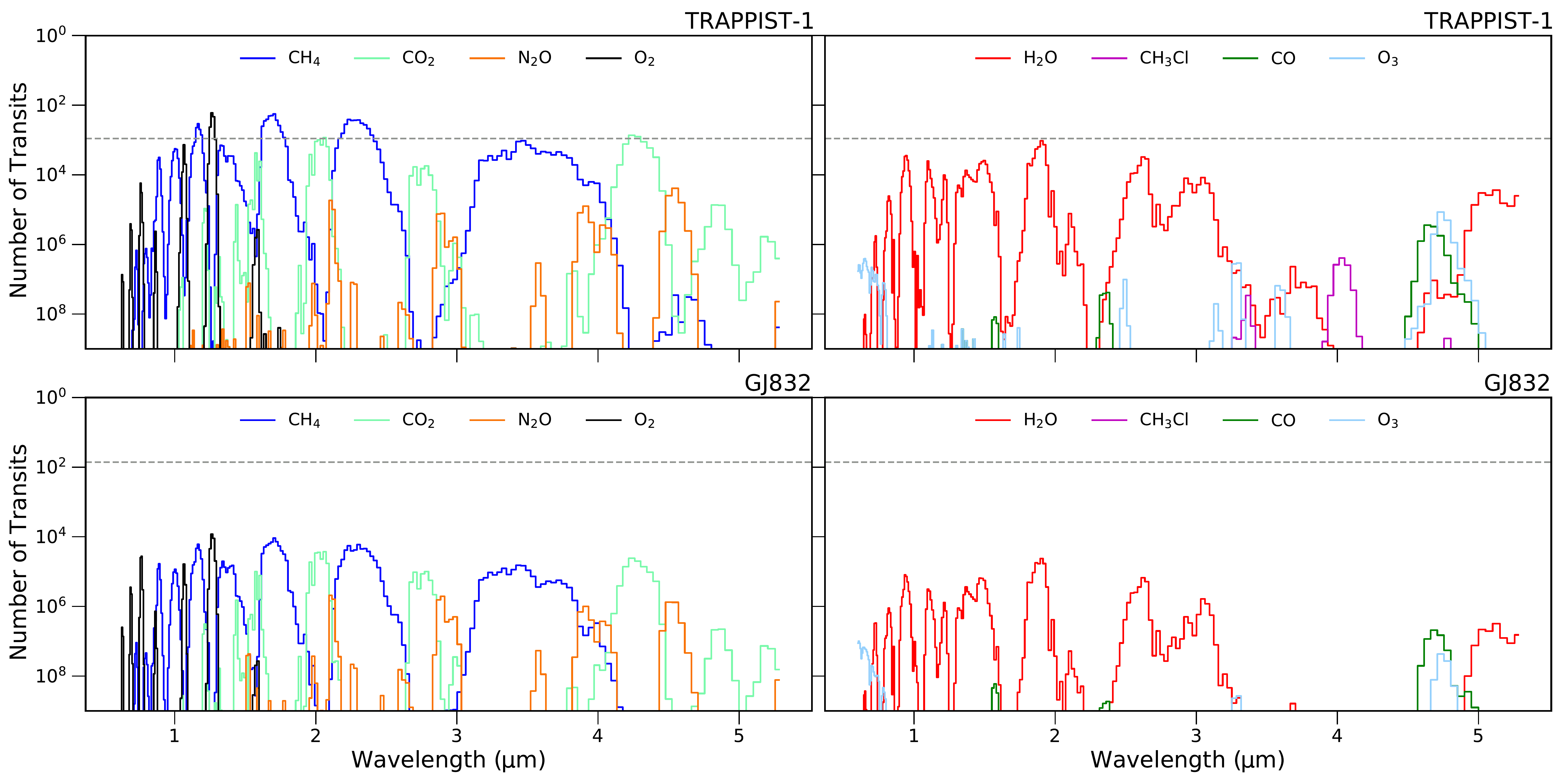}
    \caption{Number of transits required to detect the spectral impact of different gases (indicated by line color) at a signal-to-noise ratio of 5 for NIRSpec/PRISM. Scenarios are for an Earth-like planet orbiting a TRAPPIST-1-like host (top) and a GJ832-like host (bottom) at 10~pc. The grey dotted line shows the maximum number of transits that {\it JWST} could observe for each planet during a 10-year mission lifetime assuming the target is continuously visible. Gas curves peaking above the grey dotted line indicate a greater potential for detection.}
    
    \label{fig:NtVWav}
\end{figure*}

Adopting NIRSpec/PRISM for its performance and broad wavelength coverage, Figure~\ref{fig:NtVWav} explores the required number of transits to achieve a signal-to-noise ratio of 5 detection of the spectral impact (i.e., from Equation~\ref{eqn:snr_res}) for our collection of key gases. As before, we adopt TRAPPIST-1- and GJ832-like hosts at 10~pc. Here, the horizontal grey dotted line indicates the maximum number of transits that {\it JWST} could observe for each planet during in a 10-year mission lifetime. Gases whose spectral impact leads to a required number of transits that are far below the horizontal line are extremely unlikely to be detected by NIRSpec/PRISM under these conditions, while gases whose spectral impact leads to a required number of transits that peak above the horizontal line could potentially be detected.

\subsection{Wavelength-Integrated Detection Times} \label{sec:NtVDis}

\begin{figure*}
    \centering
    \includegraphics[width=\textwidth]{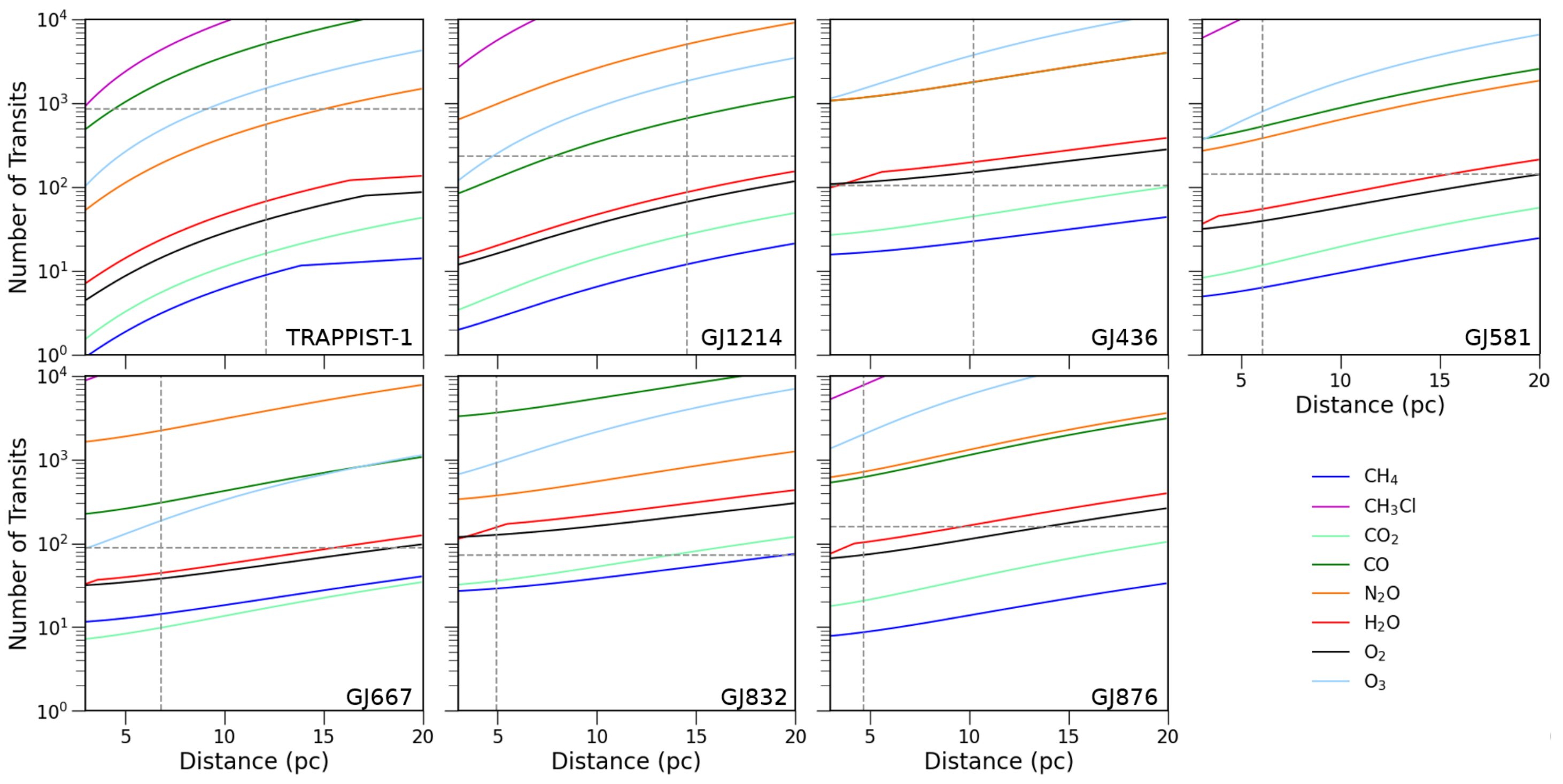}
    \caption{Required number of transits to detect the spectral impact of a given gas at a SNR of 5 for a variety of host analogs at a range of distances. Curves adopt the instrument/mode that most efficiently yields a gas detection at a given distance and the gas spectral impact is integrated over the entire range of the adopted instrument/mode. The indicated {\it JWST} limit (horizontal grey dashed) is the total number of transits {\it JWST} could observe for each star/planet combination in a 10-year mission lifetime assuming the target is continuously visible. The actual distance to the star used is shown by the vertical grey dashed line \citep[distances from][]{wunderlichetal2019}}.
    \label{fig:Predics}
\end{figure*}

Although the capabilities of {\it JWST} to detect the spectral impact of a gas at a single wavelength for our modeled scenarios appear overly challenging (Section~\ref{sec:NtVWav}), we can integrate the spectral impact of a gas over all instrument-relevant wavelengths to reduce the required number of observed transits for detection (via Equation~\ref{eqn:snr_int}).
Figure~\ref{fig:Predics} shows the number of transits required to detect each gas as a function of distance to a given stellar host analog. Signatures were integrated over wavelength, and the required number of transits for a SNR of 5 species detection at each distance point was selected as the minimum from the most promising instruments and observing modes identified in Section~\ref{sec:NtVWav} (NIRSpec/PRISM, NIRSpec high-resolution, NIRCam/F322W2, and NIRISS/SOSS).


Briefly, NIRSpec/G140H most efficiently detects O$_{2}$ for our simulated cases, independent of host or distance. For CH$_{3}$Cl, CO, CO$_2$, O$_{3}$, and N$_{2}$O, NIRSpec/G395H yields detections in the fewest number of transits, largely independent of distance and host. Similarly, NIRSpec/G235H most efficiently detects H$_2$O, largely independent of distance and host (with the exception of small distances, in which MIRI medium-resolution dominates for H$_{2}$O). NIRSpec/G235H was also shown to most efficiently detect CH$_4$. A key exception is the TRAPPIST-1 analog, whose relatively low luminosity means that NIRSpec/PRISM can be used to more efficiently detect H$_2$O and CH$_{4}$ at distances beyond 16~pc and 14~pc respectively. Furthermore, although considered here, NIRCam/F322W2 and NIRISS/SOSS were not found to be the best-case instrument for any star/gas/distance combination. A caveat to this is that in this study we only consider a small range of M type stars and terrestrial planets, for other targets (especially brighter targets) NIRCam or NIRISS may be more favorable. Finally, note that a particular instrument and observing mode providing the most efficient detection of a given species does not necessarily imply that this detection can be obtained with a reasonable investment of observing time.

%
\section{Discussion} \label{sec:discuss}
%

While Earth-sized planets have been detected in the habitable zone of M dwarf hosts, it is not yet known how similar or dissimilar these worlds may be to Earth. Nevertheless, the collection of simulated planets in Figure~\ref{fig:atmos} indicates potential atmospheric compositions for Earth-like worlds orbiting a suite of M dwarf hosts. These simulated worlds have the potential to be analogs for already-known exoplanets (e.g., TRAPPIST-1 d or e) or exoplanets that may soon be discovered, and provide an opportunity to understand the capabilities of {\it JWST} for studying the atmospheres of transiting exo-Earths.

The {\it JWST} instrument simulator developed in this work is designed to be straightforward to use and understand. Critically, this new tool compares favorably to existing tools \citep[e.g.,][]{wunderlichetal2019}, with differences in computed SNRs typically well within 10\%. Of course, the true performance of {\it JWST} and its instruments will become clear only after launch. As has been the case for the many great observatories that precede {\it JWST}, instrument capabilities may be found to improve over the duration of the mission as systematic sources of error become better understood \citep[e.g.,][]{demingetal2015}. The detectability analysis here does not include any assumed systematic ``noise floor,'' which is controversial \citep[e.g.,][]{greeneetal2016,schlawinetal2020a,schlawinetal2020b} and would serve to dramatically increase requisite exposure times for weak features that fall below the $\sim\!10$~ppm level.

When investigating the requisite number of transit observations to detect the spectrally-resolved impact of methane for a transiting exo-Earth across a wide variety of instrument modes (Figures~\ref{JWSTPerformanceTrap} and \ref{JWSTPerformance832}), NIRSpec, NIRISS, and NIRCam perform comparably, although the NIRSpec delivers detections with the smallest number of observed transits. Of course, the instrument/mode that provides the most efficient path to detecting a given species depends on a large number of factors, including host star spectral type and distance, species type and concentration, and other atmospheric parameters (such as temperature). Even for a stronger-signature scenario (i.e., the Earth analog around TRAPPIST-1 in Figure~\ref{JWSTPerformanceTrap}), spectrally-resolved methane detections require over 100 observed transits, and this remains true for all of CO$_2$, N$_2$O, O$_2$, H$_2$O, CH$_3$Cl, CO, and O$_3$ when highlighting NIRSpec/PRISM (Figure~\ref{fig:NtVWav}). Such a large number of observations is likely unrealistic for a community tool like {\it JWST}, and points to the importance of considering the spectral impact of a gas integrated across the entire wavelength range of an instrument/mode.

\begin{figure*}
    \centering
    \includegraphics[width=\textwidth]{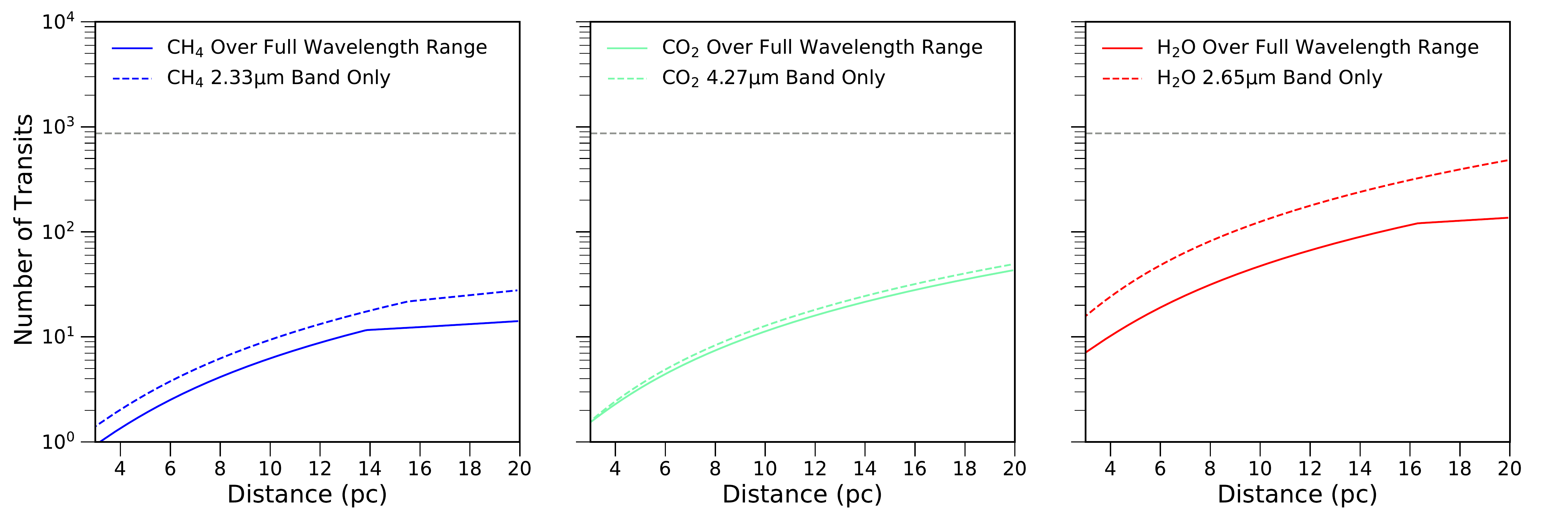}
    \caption{Required number of transits to detect the spectral impact of  CH$_{4}$ (left), CO$_{2}$ (middle), and H$_{2}$O (right) when integrating the spectral impact over the entire wavelength range of an instrument/mode (solid) and when considering only the narrow wavelength range with the strongest contribution from that gas (dotted). Only a TRAPPIST-1 analog is considered here. For the single wavelength cases, we consider the 2.3$~\upmu$m band for CH$_{4}$, the 4.3$~\upmu$m band for CO$_{2}$, and the 2.7$~\upmu$m band for H$_{2}$O. For every individual wavelength case, NIRSpec high-resolution was used and the disperser changed based on the wavelength range needed for the given gas. The grey dashed line shows the total number of transits {\it JWST} could observe for an Earth-like planet orbiting a TRAPPIST-1-like host in a 10-year mission lifetime assuming the target is continuously visible.}
    \label{fig:Comparison}
\end{figure*}

Gas detectability results where signatures are integrated over the entire wavelength range of an instrument/mode (Figure~\ref{fig:Predics}) are significantly more promising than spectrally-resolved detections. In order to highlight this, Figure~\ref{fig:Comparison} shows the required number of transits to detection for CH$_{4}$, CO$_{2}$, and H$_{2}$O when integrating spectral impact over the best case instrument's full wavelength (solid lines) and when considering only the narrow wavelength range at which each gas contributes most prominently (dotted lines). Note, the best case instrument may change at different distances (as in Figure~\ref{fig:Predics}). For the single wavelength cases, we consider  2.33$\pm{0.15}$ $\upmu$m for CH$_{4}$, 4.27$\pm{0.11}$ $\upmu$m band for CO$_{2}$, and 2.65$\pm{0.18}$ $\upmu$m  for H$_{2}$O \citep{wunderlichetal2019,robinson2019earth}.
For all three gases detection times are improved when considering an instrument's full wavelength range and the amount by which detection times are improved depends on how much each gas absorbs at other wavelengths (besides the most prominent). For example, besides at 4.27~$\upmu$m, carbon dioxide does not strongly absorb at other wavelengths where the star has bright illuminating flux, thus detectability is not greatly improved when integrating over all other wavelengths. Conversely, water absorbs at a wide array of wavelengths and, thus, detection is greatly improved by considering an instrument's full wavelength range. Additionally, due to saturation effects at small distances, the amount by which detection is improved increases as distance from the system increases.
For the Earth analogs considered here and the TRAPPIST-1-like host, methane could be detected in under 10 transits at distances within about 14~pc, carbon dioxide could be detected in under 20 transits at distances within about 13~pc, and molecular oxygen could be detected in under 40 transits at distances within about 12~pc.
Methane and carbon dioxide may also be detectable at a signal-to-noise of 5 for some mid-M dwarfs. For example, these two gases could be detected in under 10 and 20 transits, respectively for a GJ1214 analog out to distances of roughly 13 pc. Molecular oxygen and water may also be detectable for some mid-M dwarfs, but will be significantly more challenging. Other gases require rather unrealistic observational investments, except for very-nearby mid/late-M dwarf cases. Additionally, systematic noise could make these gases more difficult to detect and will impact the value gained by integrating over the entire wavelength range of an instrument/mode.

The {\it JWST} exo-Earth atmospheric characterization predictions presented here tend to fall intermediate to previous results. Emphasizing TRAPPIST-1 (likely the most widely-studied {\it JWST} case in the literature for exo-Earth scenarios), our results indicate that, generally, more transit observations are required for gas detections than in \citet{wunderlichetal2019}. This difference can largely be attributed to how SNRs are computed\,---\,\citet{wunderlichetal2019} compare the planet to an atmosphereless case as opposed to a case with just a single gas removed (as in the results here), thereby finding larger gas signature SNRs per transit. The impact of these different approaches can be significant, especially for the 4.3~$\upmu$m band of CO$_2$ which is located near many other strong bands from different gases (Figure~\ref{SNRCompare}). Additionally, differences in the underlying host stellar spectrum and photochemical model impact detectability, where the smaller ozone concentrations adopted here also contribute to the larger numbers of required transits as compared to \citet{wunderlichetal2019}. Note that the 30 transits required for ozone detection found by \citet{barstow&irwin2016} for TRAPPIST-1d is also impacted by photochemical assumptions, as the Earth-like ozone profile adopted in \citet{barstow&irwin2016} has concentrations about 10 times larger than the TRAPPIST-1 case in \citet{wunderlichetal2019} \citep[for a discussion of ozone detectability for TRAPPIST-1 planets, see also][]{krissansentottonetal2018b}. When comparing our work to these similar studies, the importance of understanding the high-energy radiation environment for a planet becomes apparent as the atmospheric model used (which is sensitive to the host stellar spectrum)  has a clear and significant impact on detection times.


Scaling reported cloud-free detection SNRs for TRAPPIST-1e from \citet{pidhorodetskaetal2020}, these authors found roughly 20, 170, 830, and 130 transits to detect CO$_2$, CH$_4$, O$_2$, and O$_3$ at an SNR of 5, respectively. Note that \citet{pidhorodetskaetal2020} use the difference between the core of a gas feature and the continuum on either side of this feature to determine the spectral impact of a given species. This approach likely results in the slightly larger number (roughly $2\times$) of required transits for detecting O$_2$. The overall larger nitrous oxide concentrations reduces the spectral impact of CO$_2$ in our work, resulting in a markedly larger number of requisite observed transits for detection. Differences in predicted methane and ozone concentrations make these gases easier and harder to detect in our work, respectively, as compared to \citet{pidhorodetskaetal2020}.

\citet{macdonald&cowan2019} discuss the detectability of key atmospheric features for an assumed Earth-like TRAPPIST-1e using an empirically-derived transit spectrum of Earth. Adopting a noise model for NIRSpec and MIRI, and by comparing the depths of the single strongest feature for a given gas to the nearby continuum, \citet{macdonald&cowan2019} find that roughly three transits are required to detect CO$_2$ and that 130 transits would be required to detect H$_2$O. (Predictions are also provided for CH$_4$ and O$_3$, but comparing to these predictions is not meaningful as the atmospheric models adopted here have markedly different abundance profiles for these gases as compared to the actual Earth.) Our calculations indicate that roughly 15 transits would be required for an equivalent CO$_2$ detection, which is larger than the estimates from \citet{macdonald&cowan2019} and is potentially due to our isolation of the transit depth signal due solely to CO$_2$ rather than an in-band versus out-of-band approach. For H$_2$O, we find that only about 70 transits are required for an equivalent detection. This number is slightly smaller than the estimates from \citet{macdonald&cowan2019}, and the decrease is likely due to our combining gas absorption signals across all observable bands.

Finally, one key caveat for the detectability results presented above is that these are derived from cloud-free climate and transit spectrum models. Especially for water vapor, whose concentrations tend to be highest in the near-surface atmosphere beneath any clouds, cloud extinction in transit spectra can strongly limit detectability \citet{komaceketal2020,suissaetal2020}. The impacts of clouds can be less severe for well-mixed gases or gases whose concentration peak in the upper atmosphere, as the cores of associated absorption bands can probe altitudes above condensate clouds. For example, \citet{krissansentottonetal2018b} saw only minor impacts from clouds on CO$_2$ and CH$_4$ detections for an Archean Earth-like TRAPPIST-1e. Similarly, \citet{pidhorodetskaetal2020} see relatively minor detection SNR reductions due to clouds for CO$_2$ and O$_3$ (which have feature that saturate above the clouds), moderate reductions for CH$_4$ (whose concentrations are, again, much lower than those modeled here), and strong reductions for O$_2$ (whose relatively weak features tend to probe at/below the cloud level). Additional efforts that couple photochemical tools to cloud-simulating general circulation models are needed to more fully understand the interplay between cloud coverage and species detectability, especially when considering realistic atmospheric retrieval scenarios for {\it JWST}.

\section{Conclusions} \label{sec:conclusions}

In this paper, we explored the capabilities of {\it JWST} to detect key biosignature and habitability indicator gas species for Earth-like exoplanets orbiting seven different M dwarf-type hosts, including a TRAPPIST-1-like host. Our key results and findings are:

\begin{itemize}
    \item We developed and validated a straightforward tool for simulating exoplanet transit spectra with {\it JWST}.
    
    \item Integrating the spectral impact of a gas over the entire wavelength range of a given instrument/mode yields a requisite number of observed transits for detection that are substantially smaller than when considering detectability in a single spectral element at the core of strong absorption features. Importantly, this combination of information over the full range of applicable wavelengths is most analogous to full atmospheric retrieval analyses.
    
    \item Certain species (especially methane and carbon dioxide) could be detected with {\it JWST} in several tens of transits (or less) for mid/late-type M dwarfs hosting Earth-like transiting exoplanets at system distances of up to 10--15~pc.
    
    \item Via comparisons to previous studies, the {\it JWST} gas detection results presented here are shown to be strongly dependent on the assumed underlying atmospheric chemistry, host star spectrum, and the approach to computing detection SNRs. More detailed work is required with coupled chemistry-climate tools, and actual {\it JWST} gas concentration constraints could then help further our understanding of terrestrial planet atmospheric chemistry.
\end{itemize}

\section{Acknowledgements}

We thank an anonymous reviewer for thoughtful and constructive suggestions.
MTG would like to acknowledge funding from the Hooper Undergraduate Research Award at Northern Arizona University (NAU), the NAU/NASA Space Grant program, and the Goldwater Scholarship. TDR gratefully acknowledges support from NASA's Exoplanets Research Program (No.~80NSSC18K0349) as well as the Nexus for Exoplanet System Science and NASA Astrobiology Institute Virtual Planetary Laboratory (No.~80NSSC18K0829). FW acknowledges financial support from DFG project RA-714/7-1. SR gratefully acknowledges support from the Glasstone Fellowship. This work has made use of the MUSCLES M dwarf UV radiation field database.
%

\bibliography{main}{}
\bibliographystyle{aasjournal}

\end{document}